\documentclass{vgtc}                          

\graphicspath{{figures/}{pictures/}{images/}{./}} 

\usepackage{times}                     
\usepackage{microtype}

\usepackage{tabu}                      
\usepackage{booktabs}                  
\usepackage{lipsum}                    
\usepackage{mwe}                       

\usepackage{mathptmx}                  
\usepackage{ccicons}

\usepackage{balance}

\newcommand{\paperCountInMMXXIV}{124}

\newcommand{\humanResearchInMMXXIV}{93}

\newcommand{\humanResearchPercentageInMMXXIV}{75.0}
\newcommand{\humanResearchFullyCorrectInMMXXIV}{4}
\newcommand{\humanResearchFullyCorrectWordInMMXXIV}{four}
\newcommand{\humanResearchFullyCorrectPercentageInMMXXIV}{4.3}

\newcommand{\humanResearchMostlyCorrectInMMXXIV}{14}

\newcommand{\humanResearchLargelyinorrectInMMXXIV}{79}
\newcommand{\humanResearchLargelyinorrectWordInMMXXIV}{seventy-nine}
\newcommand{\humanResearchMostlyCorrectPercentageInMMXXIV}{15.1}

\newcommand{\paperCountInMMXXV}{131}

\newcommand{\humanResearchInMMXXV}{96}

\newcommand{\humanResearchPercentageInMMXXV}{73.3}
\newcommand{\humanResearchFullyCorrectInMMXXV}{2}
\newcommand{\humanResearchFullyCorrectWordInMMXXV}{two}
\newcommand{\humanResearchFullyCorrectPercentageInMMXXV}{2.1}

\newcommand{\humanResearchMostlyCorrectInMMXXV}{12}

\newcommand{\humanResearchLargelyinorrectInMMXXV}{84}
\newcommand{\humanResearchLargelyinorrectWordInMMXXV}{eighty-four}
\newcommand{\humanResearchMostlyCorrectPercentageInMMXXV}{12.5}

\newcommand{\paperCountOverall}{255}

\newcommand{\humanResearchOverall}{189}

\newcommand{\humanResearchPercentage}{74.1}
\newcommand{\humanResearchFullyCorrect}{6}

\newcommand{\humanResearchFullyCorrectPercentage}{3.2}

\newcommand{\humanResearchMostlyCorrect}{26}
\newcommand{\humanResearchLargelyinorrect}{163}
\newcommand{\humanResearchLargelyinorrectWord}{one hundred and sixty-three}

\newcommand{\humanResearchMostlyCorrectPercentage}{13.8}
\newcommand{\percentageLiableBecauseNotMostlyCorrect}{63.9}

\newcommand{\new}[1]{\textcolor{RoyalBlue}{#1}}
\renewcommand{\new}[1]{#1}

\newcommand{\eg}{e.\,g.}
\newcommand{\ie}{i.\,e.}
\newcommand{\osfrepo}{\href{https://osf.io/3hscn/}{\texttt{osf\discretionary{}{.}{.}io\discretionary{/}{}{/}3hscn}}}
\renewenvironment{quote}%
	{\list{}{\rightmargin=\parindent \leftmargin=\parindent}%
	\item\relax}%
	{\endlist}

\onlineid{0}

\vgtccategory{Research}

\title{More Than 63\% of IEEE VIS Research Liable to be Retracted?!\\ Ethics Approval Statements Protect Participants (and Researchers!)}

\author{Lonni Besançon\,\href{https://orcid.org/0000-0002-7207-1276}{\includegraphics[height=.8em]{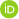}}\hspace{1pt}\thanks{e-mail: lonni.besancon@gmail.com}\\ %
        \scriptsize\parbox{0.4\textwidth}{\centering Linköping University, Sweden}%
\and Tobias Isenberg\,\href{https://orcid.org/0000-0001-7953-8644}{\includegraphics[height=.8em]{orcid.pdf}}\hspace{1pt}\thanks{e-mail: given\_name.family\_name@inria.fr}\\ %
     \scriptsize\parbox{0.4\textwidth}{\centering Université Paris-Saclay, CNRS, Inria, LISN, France}}
\abstract{%
    We analyzed the ethics reporting in \paperCountOverall{} IEEE VIS papers from 2024 and 2025, as published in TVCG. This analysis arose from our experience as readers and reviewers of IEEE VIS papers that such reporting is frequently incomplete or missing, as well as from investigations in which we ourselves had to answer challenges regarding ethics approval in our own work. Visualization research naturally often involves human participants, yet ethics approval and informed-consent procedures are not always explicitly reported. In our corpus, \humanResearchOverall{} papers (\humanResearchPercentage\%) reported on work involving human participants. Only \humanResearchFullyCorrect{} of them (\humanResearchFullyCorrectPercentage\%) reported to have obtained ethics approval, an approval identifier, and having received informed consent from the participants, while \humanResearchMostlyCorrect{} (\humanResearchMostlyCorrectPercentage\%) reported at least ethics approval and informed consent. These omissions do not imply that the empirical work was unethical or lacked approval. They rather show that current reporting practices make ethical approval difficult to assess. Beyond our alarming---yet true---pa\-per title, we wish to raise awareness on how authors themselves may eventually be at risk for not properly reporting ethics. We argue that VIS should adopt clearer and more standardized ethics-re\-por\-ting practices to protect participants, authors, reviewers, and editors.} 


\begin{document}


\firstsection{Introduction}

\maketitle

Visualization research, at least since the inclusion of the historical InfoVis and VAST, has almost always relied on user experiments to validate systems, algorithms, theories, or hypothesis \cite{Lam_2012,Isenberg_2013,Lin}. This fact is not surprising as visualization is often considered as being a subfield of Human-Computer Interaction, itself particularly relying on human-subject studies \cite{Caine}. Our practices of evaluation are constantly evolving and are being questioned as evidence by the numerous BELIV workshops that we have held so far and the surprisingly large amount of methodological papers looking at our evaluation or reporting practices within our outside of BELIV. 

In the last couple of years, for instance, our field has reflected upon and mostly adapted, based on discussions, position papers, and analysis of our own past practices, new or more thorough study reporting strategies. Practices from other fields have thus been adopted or discussed such that we now consider preregistration \cite{prereg_beliv}, data and code sharing when relevant and appropriate \cite{haroz:hal-01947432}, different statistical analyses \cite{cockburn:hal-02907143,dragicevic:hal-01377894}, new transparency and rigor guidelines for design studies \cite{Meyer_Rigor,Rogers}, checklists for evaluating evaluation papers \cite{Crisan}, reflections on the reproducibility of visualization research \cite{Isenberg,Kosara,Valdez}, or even full frameworks to ensure scrutanizability of study designs such as ReVISit \cite{Cutler,Nobre}. In parallel, our reporting in papers has also evolved with our template for IEEE VIS now allowing to put references, links to the supplemental material, figure credits, and acknowledgments sections on the last two pages of the paper as well as allowing to submit appendices directly after the reference section and after the 11 pages of the paper to facilitate referencing of materials in the main paper. 

For the 2026 IEEE VIS conference, another major change was made for papers submitted to IEEE VIS focusing on the reporting of work with human participants.\footnote{\new{For the definition we adopt in this paper for research with human participants please see \autoref{sec:definition}.}} The paper submission guidelines page currently reads:\footnote{See \href{https://ieeevis.org/year/2026/info/call-participation/paper-submission-guidelines/}{\texttt{ieeevis\discretionary{}{.}{.}org\discretionary{/}{}{/}year\discretionary{/}{}{/}2026\discretionary{/}{}{/}info\discretionary{/}{}{/}call\discretionary{}{-}{-}participation\discretionary{/}{}{/}paper\discretionary{}{-}{-}submission\discretionary{}{-}{-}guidelines}}.}
\begin{quote}
    \emph{The IEEE Publication Services and Products Board Operations Manual, which governs the publication in TVCG, clearly specifies that ``Authors of articles reporting on research involving human subjects or animals, including but extending beyond medical research, shall include a statement in the article that the research was performed under the oversight of an institutional review board or equivalent local/regional body, including the official name of the IRB/ethics committee, or include an explanation as to why such a review was not conducted. For research involving human subjects, authors shall also report that consent from the human subjects in the research was obtained or explain why consent was not obtained.''}
\end{quote}

This change may, at first, appear like another layer of bureaucratic or administrative information that would further increase the burden on paper authors. The reason for this change, however, is to protect our participants and our own community as authors. Indeed, such requirements already appear, as cited above, for the journal publishing our proceedings, IEEE TVCG, and we should respect it to make sure our papers are aligned with the journal and publisher's policies. 

Through this paper we propose to explain why such a requirement is important even beyond IEEE and TVCG policies and analyze current reporting practices from two years of IEEE VIS papers (VIS 2024 and VIS 2025). We, finally, close the paper with reflections on the process along with ideas for improvements.

\new{Overall, our contributions with this paper are the following: (1) an analysis of the ethics approval and informed consent statement in the past two years of VIS papers highlighting that \percentageLiableBecauseNotMostlyCorrect{}\% of all papers published at IEEE VIS are violating the IEEE's own requirements for publishing research that involves human participants; (2) a contextualization of these findings and discussion of why ethical statement are important to protect researchers and participants; and (3) suggestions for future reporting of this information in our papers.}

\section{Background}
\label{sec:background}

Regulation of research on human participants became a priority issue after the horrors of World War 2 research \cite{Oak}. Further unethical studies in  the following decades (\eg, the Tuskegee Syphilis Study \cite{corbie1999continuing} or the thalidomide adverse effects scandal \cite{waggoner2022clinical}) have reinforced the attention given to the problem. Clinical research is particularly surveyed and international declarations (\eg, the declaration of Helsinki \cite{goodyear2007declaration,williams2008declaration}) tend to govern best practices while national regulations impose strict practices to protect participants. Publishers have also signed the declaration of Helsinki and COPE (the committee on publication ethics) recommends following it. National regulations are all very different and difficult to summarize, even for the case of clinical research.

Clinical research, on the one hand, is particular because most of it goes beyond the notion of ``minimal risk'' and thus requires that strict procedures and notions are followed and that they are justified. The research we conduct in HCI and visualization rarely goes, on the other hand, beyond the notion of minimal risk. In most cases, we ask participants to perform tasks that they would perform on a daily basis (clicking on a computer, manipulating a smartphone, or wearing a VR/AR headset). These studies are thus very simple (ethically) and do not involve any risk or invasive procedures that clinical research studies more often present. National regulations on the matter thus vary even more than for clinical research. Until fairly recently, for instance, Inria---where Lonni was a PhD student and Tobias still works---did not have specific processes or national regulations on the matter to follow. 

\new{However, the lack of guidelines could have cost us both (and our co-authors) a lot.\footnote{\new{We provide a full account of this story in \autoref{sec:personal}.}} During the pandemic, Lonni became particularly involved with post-publication scrutiny, finding and calling out problematic scientific practices (\eg peer-review conducted in a single day \cite{besanccon2021open}) or papers (\eg \cite{meyerowitz2021impact}). A particular focus of his work at the time was on papers published by a single research institute which cumulated several questionable practices: single-day peer review combined with editorial conflict of interest \cite{besanccon2021open}, inconsistent data across re-analysis from the same authors \cite{barraud2023article}, and, most of all, issues around ethical approval \cite{barraud2023article}. This situation prompted Lonni and some of his colleagues to systematically investigate papers from this institute for ethical issues \cite{frank2023raising} only to find numerous missing, unclear, or repeatedly reused ethics approval statements, including one approval number reported in more than 248 papers involving different populations, samples, and countries \cite{frank2023raising}. This investigation has, thus far, led to more than 70 retractions and the identification of ethical or legal concerns in 853 articles of the same institute \cite{frank2026followup}.}

\new{Following this work, similar scrutiny was directed at some of Lonni's visualization and HCI publications, prompting investigations by both publishers and institutions. Some of these publications did not include proper ethics approval number or exemption information despite the requirement to do so on the side of the publisher. These investigations eventually cleared the work: for the papers concerned, the relevant approvals or participant protections procedures either did not exist or had been followed, simply not reported. Yet, responding to them was time-consuming. It required reconstructing institutional requirements and practices that were often only documented in French and using versions of guidelines that are, in some cases, not always available online anymore. Based on this experience, first, it was good to see that post-publication scrutiny given to our papers was taken seriously by publishers \cite{Besancon_correction}, but it also allowed us to think about the practice in the field and what we could do about it. Second, it allowed us to reflect about our practices in the field and motivated us to examine systematically how ethics and consent to participate was reported in VIS papers.}

\section{Investigating ethics reporting in IEEE VIS papers from VIS 2024 and VIS 2025}

To thus assess the state of ethics reporting in IEEE VIS papers, we looked into the papers published at the 2024 and 2025 conferences. We downloaded the respective special issues from IEEE TVCG (the journal publishing the proceedings of the conference) in 2025 and 2026 respectively (\ie, the January issues in the year following the conference)\footnote{In our discussion in the remainder of this paper we refer to the year of the conference rather than the publishing year of the TVCG special issue.} and followed the method we outline below.

\subsection{Method}

For each paper, one coder assessed whether the article reported the following information:

\begin{itemize}
\item \textbf{human participants:} whether the article reported on work that involves human participants---including \emph{\textbf{all forms of feedback or data from people other than the authors}}
\item \textbf{ethics approval:} whether the article reported that the work had been approved by an ethics committee (or similar) or had been granted an exemption;
\item \textbf{ethics approval number:} whether the article reported an ethics approval number or identifier;
\item \textbf{informed consent:} whether the authors reported having obtained informed consent from participants.
\end{itemize}

Because articles in our field often report multiple user studies within a single paper, we also coded \emph{partial} responses for each of the above categories when the information appeared to apply to only some, but not all, of the reported studies.

Finally, we checked whether an LLM, ChatGPT 5.5 Thinking, can answer the same questions automatically for all articles. For the IEEE VIS 2024 corpus, we did the LLM-based analysis alongside the manual coding to identify mismatches during the coding process and resolve them by re-reading the relevant article. For the IEEE VIS 2025 corpus, we conducted the LLM-based assessment after the manual coding. We used the following prompt:
\begin{quote}
I want you to read through these articles in the linked PDF and check whether they have an experiment that involves human beings (even online). If they do, I want you to check if they mention having obtained ethics approval, if they mention an ethics approval number, if they mentioned having informed consent, and any needed notes to help interpret this. 
\end{quote}

We recorded our manual and the LLM-based assessments separately, and used any potential mismatches between them as prompts for additional (manual) verification of the relevant articles---which indeed sometimes lead to us updating our coding. Sometimes, for instance, the LLM found additional information in (pixel-based) images in the paper, in acknowledgments sections, or in additional material that was not part of the original downloadable paper PDF. In other cases the LLM was wrong or did not fully understand the used expressions, which we then explained in a notes/differences column in our data. We 
provide our final coding, the LLM-generated codes, our comments on the differences, and our scripts that extract the final numbers and the graphical plot as additional material.

\subsection{Results}

\begin{figure}[t]
	\centering
	\includegraphics[width=\columnwidth]{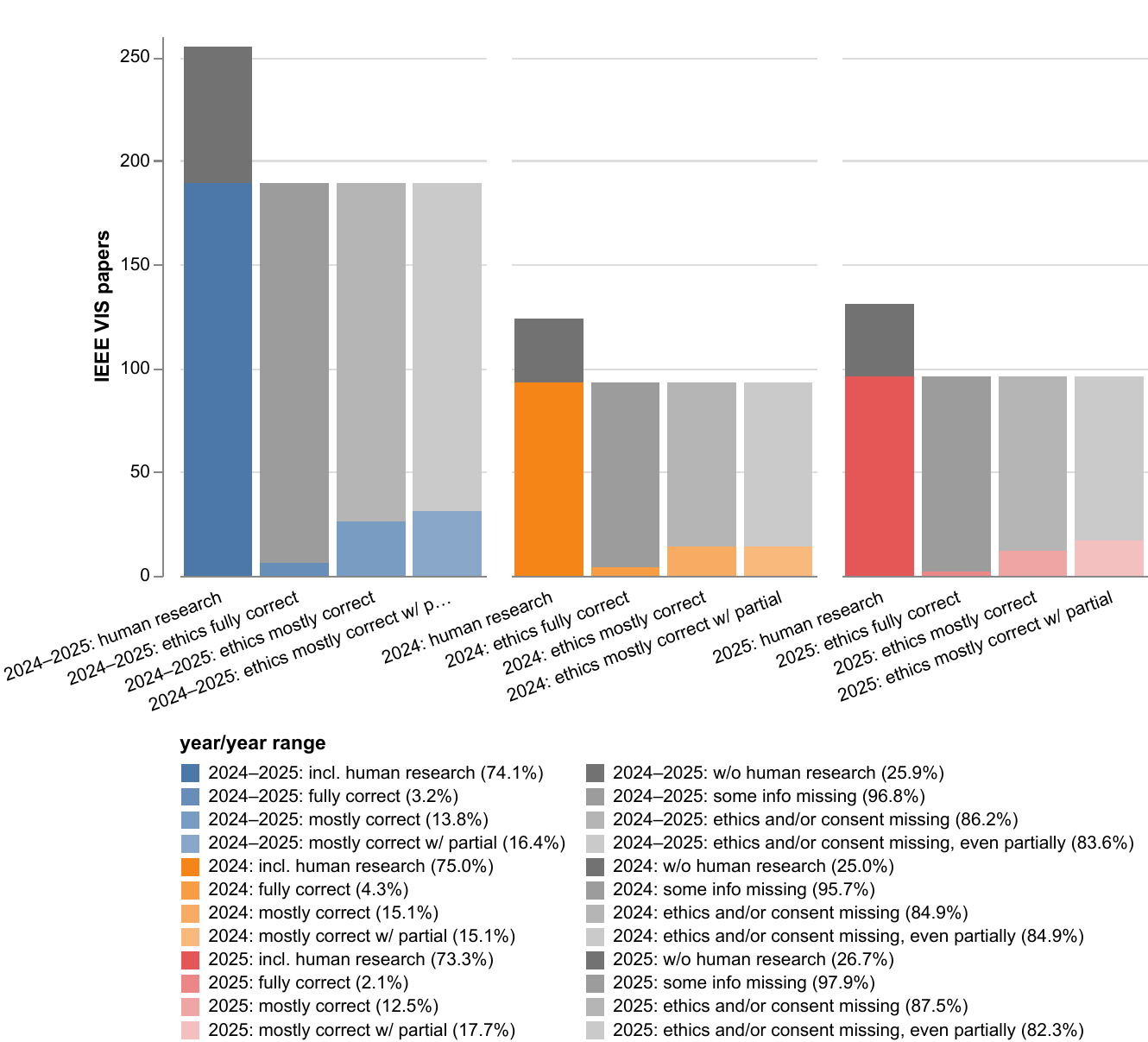}\vspace{-1ex}
	\caption{Bar chart of results. For a normalized version see \autoref{fig:ethics_aggregated-stackedbargraph-normalized}.}\vspace{-2ex}
	\label{fig:ethics_aggregated-stackedbargraph}
\end{figure}

\begin{figure}[t]
	\centering
	\includegraphics[width=\columnwidth]{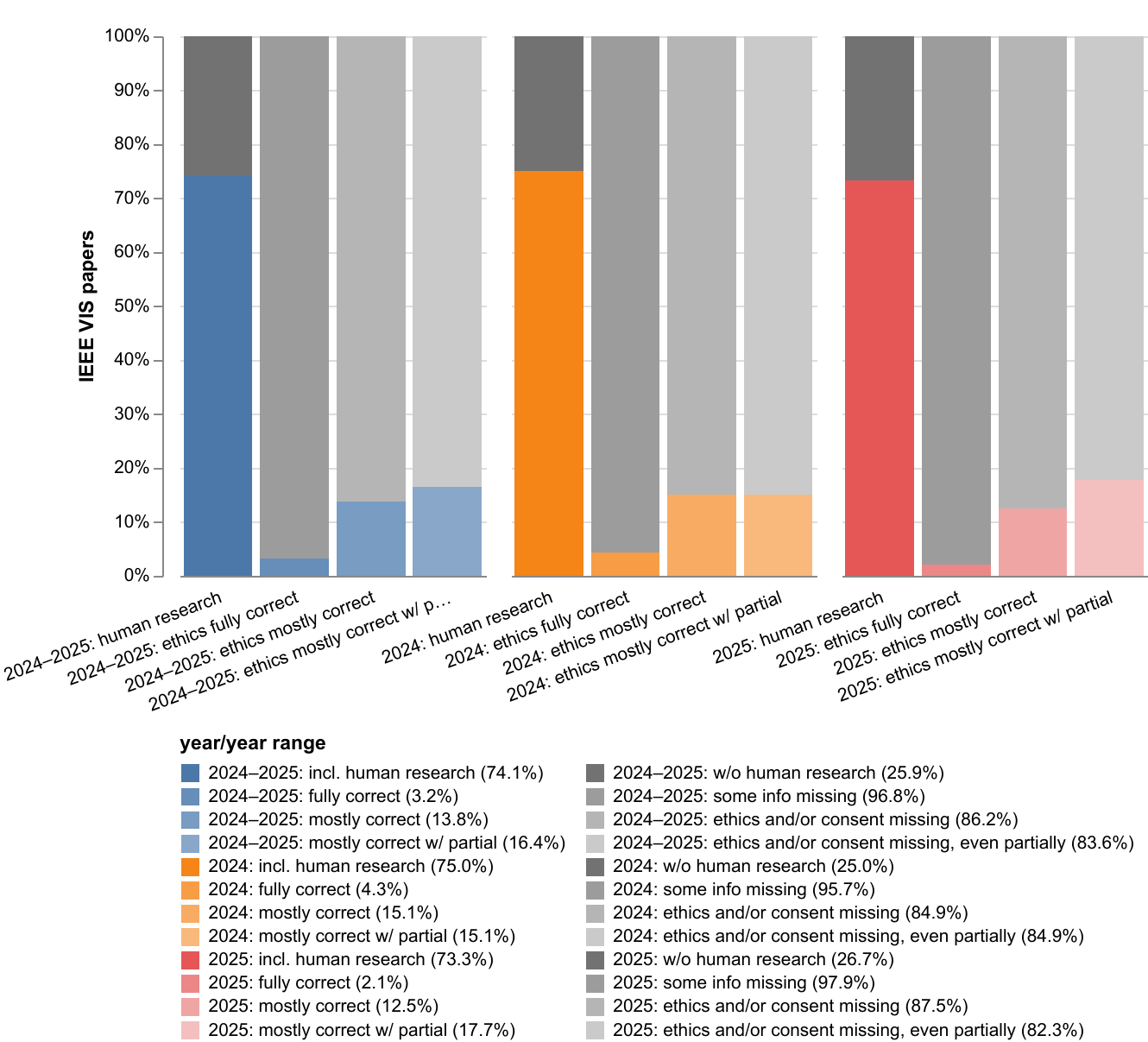}\vspace{-1ex}
	\caption{Normalized bar chart of our results. For a version with absolute numbers see \autoref{fig:ethics_aggregated-stackedbargraph}.}\vspace{-2ex}
	\label{fig:ethics_aggregated-stackedbargraph-normalized}
\end{figure}

We screened a total of \paperCountOverall{} articles with \paperCountInMMXXIV{} for IEEE VIS 2024 (Issue 1 of TVCG 2025) and \paperCountInMMXXV{} for IEEE VIS 2025 (Issue 1 of TVCG 2026). Out of these \paperCountOverall, we found that \humanResearchOverall{} reported an experiment with or feedback from human participants who were not among the paper authors (\humanResearchPercentage\%). Per year, this is a breakdown of \humanResearchInMMXXIV{} out of \paperCountInMMXXIV\ (\humanResearchPercentageInMMXXIV\%) for IEEE VIS 2024 and \humanResearchInMMXXV{} out of \paperCountInMMXXV{} (\humanResearchPercentageInMMXXV\%) for IEEE VIS 2025 (see the bar charts in \hyperref[fig:ethics_aggregated-stackedbargraph]{Figures}~\ref{fig:ethics_aggregated-stackedbargraph} and \ref{fig:ethics_aggregated-stackedbargraph-normalized}).

Now considering the reporting of ethics approval and informed consent, we found the following. If we do not require the reporting of the specific ethics approval number, then a total of \humanResearchMostlyCorrect{} papers (\humanResearchMostlyCorrectPercentage\% of all papers with human participants) with \humanResearchMostlyCorrectInMMXXIV{} papers for VIS 2024 (\humanResearchMostlyCorrectPercentageInMMXXIV\%) and \humanResearchMostlyCorrectInMMXXV{} papers for VIS 2025 (\humanResearchMostlyCorrectPercentageInMMXXV\%) reported the ethical aspects of the work with human participants satisfactorily.\footnote{We coded additional papers as ``partially'' fulfilling the requirements, such as partially reporting ethics review, partially naming an approval number, or partially reporting to have obtained informed consent. While such partial fulfillment does not make a difference for 2024, it changes the results slighly for 2025 as we show in \hyperref[fig:ethics_aggregated-stackedbargraph]{Figures}~\ref{fig:ethics_aggregated-stackedbargraph} and \ref{fig:ethics_aggregated-stackedbargraph-normalized}.}

Of note, only \humanResearchFullyCorrectWordInMMXXIV{} (\humanResearchFullyCorrectInMMXXIV) papers \cite{cabouat:hal-04665390,Han,Leon,Zhao} from IEEE VIS 2024 (\humanResearchFullyCorrectPercentageInMMXXIV\% of all papers with human participants) and \humanResearchFullyCorrectWordInMMXXV{} (\humanResearchFullyCorrectInMMXXV) papers from IEEE VIS 2025 \cite{Ferron_2025,He_2026} (\humanResearchFullyCorrectPercentageInMMXXV\%) reported ethics approval, the ethics approval number, and having obtained informed consent from participants (\ie, a total of \humanResearchFullyCorrect{} papers for both years combined or \humanResearchFullyCorrectPercentage\% of all papers with human participants). Perhaps unsurprisingly, three (3) of the \humanResearchFullyCorrectWordInMMXXIV{} (\humanResearchFullyCorrectInMMXXIV) papers from VIS 2024 and all of the \humanResearchFullyCorrectWordInMMXXV{} (\humanResearchFullyCorrectInMMXXV) from  VIS 2025 \cite{Ferron_2025,He_2026} involve authors from (or formerly from) AVIZ, Inria, who have co-authored with us and have, as such, also been under scrutiny in the past (see \autoref{sec:personal}).

What these results also show is, unfortunately, that \percentageLiableBecauseNotMostlyCorrect{}\% of all papers published at IEEE VIS in the past two years---\textbf{\emph{almost two thirds}}---are violating the IEEE's own requirements for publishing research that involves human participants\footnote{\new{The IEEE Publication Services and Products Board Operations Manual asks for reporting ethics approval and informed consent at least since \href{https://web.archive.org/web/20230308023843/https://pspb.ieee.org/images/files/PSPB/opsmanual.pdf}{its January 2023 version}---the earliest archived by \href{https://archive.org/}{\textsc{The Internet Archive}}.}}\textsuperscript{,}\footnote{Including, unfortunately, some papers of our own.} and is thus, technically similarly liable to attempts by others to get investigated and potentially retracted due to a violation of publishing guidelines.

\section{Discussion}

We now summarize and contextualize our findings, discuss why they are important to consider for future IEEE VIS conferences and TVCG submissions, and discuss editorial processes and how they could evolve to support ethics reporting better.

\subsection{Human-based evaluation in visualization research}

First, 
it is important to contextualize our results with respect to past work that analyzed VIS or HCI papers based on their use of human participants in the evaluation of research work. Such prior work provides us with related, but not directly identical, benchmarks for the prevalence of studies and evaluations in HCI and visualization. At CHI, approximately 90\% of papers have been reported to include a study or evaluation \cite{Barkhuus,besancon:hal-03342756,Caine}. In visualization, Isenberg et al. \cite{Isenberg_2013} found that 569 of 581 (97\%) IEEE Visualization papers (\ie, excluding the former InfoVis and VAST conferences) in the years 1997, 2000, 2003, and 2006--2012 included at least one form of evaluation, although not all of these evaluations involved human participants. Differences existed in the different sub-communities at the time, with the InfoVis conference (based also on the numbers from Lam et al.'s \cite{Lam_2012} initial benchmark) often having 70--100\% papers with human participants, while the original Vis conference started low (10--30\%) and had increased to more than 50\% in 2012. More recent work by Lin et al.\ \cite{Lin} has estimated that that 63.55\% of the papers from IEEE VIS 2023 and 2024 that they subjectively judged to address abstract data visualization they found to include at least one quantitative evaluation. Our percentage of \humanResearchPercentage\% papers with work that generally involves human participants are similar to these numbers (but for all of the IEEE VIS proceedings in 2024 and 2025), and they indicate that human-centered evaluation has long been and continues to be a substantial component of VIS research.

\subsection{Lack of ethics reporting does not equate lack of ethics approval}
\label{sec:just-not-reported}

It is important to highlight that a lack of ethics approval reporting does not equate, in any way, a lack of ethics approval or any kind of scientific misconduct. Unlike the \new{cases in medical/microbiological research we had mentioned (\autoref{sec:background}, \autoref{sec:personal}), which potentially concern} invasive studies \cite{frank2023raising}, studies in HCI and visualization are often considered to involve a minimal risks and have, in some countries, even a blanket ethics approval. Our own cases have shown that a lack of reporting in our papers (of ethics approval numbers, of an exemption from a review, or of lack of obligations to apply under national regulatory texts) did not imply a lack of either ethics approval or misconduct. 
Despite our intentionally alarmist title, our results should thus not be taken as a message that VIS research has been unethical or not following international standards on research done on human beings. They simply demonstrate that---thus far---our reporting strategies have, on the one hand, failed to accurately account for our processes and the laws that govern our research and, on the other hand, follow the submission guidelines of our publishers \new{(IEEE, Wiley, Elsevier, ACM).\footnote{\new{Find excerpts of the respective regulations in \autoref{sec:regulations}.}}} This situation is in direct contrast with the results of Lonni's previous investigations where publishers have noted, for instance, that \emph{``the exclusion criterion raises concerns that the subjects were recruited prospectively and the ethics approval was granted retrospectively,''} \cite{pone.0346849}; \emph{``health-safety enforcement powers suspending research at IHU-MI''} \cite{ansm2022bmrstud}; a journalist from Science confirming that the hosting university tried its best to sweep a report establishing ethical issues under the rug \cite{ogrady2024unearthed}; PLOS documenting editorial concerns that ethics approvals may have been reused beyond their apparent scope and jurists from Aix-Marseille University repeatedly defending the apparent applicability of the approvals (see, \eg, \cite{pone.0324349,pone.0346619,pone.0346941}); and the French investigations eventually pushing the government to refer the matter to the courts.\footnote{See \href{https://www.lemonde.fr/planete/article/2022/09/05/ihu-de-marseille-dirige-par-didier-raoult-les-ministres-de-la-sante-et-de-la-recherche-saisissent-la-justice_6140291_3244.html?srsltid=AfmBOop8gXsxa2gnTbxusDIakZsQxou_ZylPY5cjbCFc4r-aK56vEcWY}{\textls[-25]{\texttt{lemonde\discretionary{}{.}{.}fr\discretionary{/}{}{/}planete\discretionary{/}{}{/}article\discretionary{/}{}{/}2022\discretionary{/}{}{/}09\discretionary{/}{}{/}05\discretionary{/}{}{/}ihu\discretionary{}{-}{-}de\discretionary{}{-}{-}marse\discretionary{i}{}{i}lle\discretionary{}{-}{-}dirige\discretionary{}{-}{-}par\discretionary{}{-}{-}didier\discretionary{}{-}{-}raoult\discretionary{}{-}{-}les\discretionary{}{-}{-}ministres\discretionary{}{-}{-}de\discretionary{}{-}{-}la\discretionary{}{-}{-}sante\discretionary{}{-}{-}et\discretionary{}{-}{-}de\discretionary{}{-}{-}la\discretionary{}{-}{-}recherche\discretionary{}{-}{-}saisissent\discretionary{}{-}{-}la\discretionary{}{-}{-}justice\discretionary{\_}{}{\_}6140291\discretionary{\_}{}{\_}3244\discretionary{}{.}{.}html\discretionary{?}{}{?}srsl\discretionary{t}{}{t}id\discretionary{=}{}{=}AfmBOop8gXsxa2gnTbxusDIakZsQxou\discretionary{\_}{}{\_}ZylPY5cjbCFc4r\discretionary{}{-}{-}aK56vEcWY}}}.}

In fact, our findings are not critically departing from previous investigations into other domains. Previous research has shown that even in health research, failure to report ethics approval was not uncommon although this did not mean that ethics approval was not obtained \cite{asplund2021reporting} and the number of papers not reporting ethics approval was not particularly high while the number of papers not reporting ethics approval number or reference can be much higher \cite{wu2019reporting}. An older investigation reports potentially higher rates \cite{murphy2015reporting} (within health sciences, but in different sub-fields), pointing to the fact that either reporting practices have evolved or that the number are highly field-dependent, or both. For health-related case reports, the numbers are not so positive, with a critical under-reporting of ethics \cite{tran2024reporting,valesic2025informed}. Outside of health-related research, Mayne and Howitt \cite{mayne2014reporting} have looked at ethics reporting in early-childhood education journals and found that most papers and authors failed to report on parental consent, child consent, or ethics approval, although parental consent was reported at higher rates than the others. Similar past research in empirical software engineering research has shown that a majority of studies failed to report on ethics \cite{liebel2021ethical} while research on ethics in the domain of mining software repositories highlighted that REC/IRBs may not always have all the necessary understanding for specific fields \cite{gold2022ethics}. A lack of sufficient reporting despite the presence of guidelines was also reported in educational research \cite{lees2021variation}. Similarly, linguistics research seems to suffer from insufficient reporting, although attention to ethics is omnipresent \cite{ghanbar2026principle}. Fields using publicly available data (\eg, social media posts) would tend to rarely seek ethics approval \cite{kessler2024netnographic,stommel2021ethical}, despite very real ethical questions \cite{Fiesler,proferes2021studying}.\footnote{As also highlighted in the BELIV 2022 keynote by Casey Fiesler: ``Data is People'' (\href{https://beliv-workshop.github.io/2022/keynote.html}{\texttt{beliv-workshop.github.io/2022/keynote.html}}).} Finally, on a similar topic, psychological research into misinformation also seems to un\-der-re\-port on the specific ethical practices that they should put in place for their experiment (to safeguard participants against the danger of misinformation) \cite{greene2023best}, although here again the authors highlight that their study can only look at reporting practices without actually judging whether best ethical practices were followed.  Our results are thus compatible with previous results outside of health-re\-la\-ted research, and slightly compatible with some of practices in health-re\-la\-ted research. In all this work, however, the authors are careful not to point to misconduct but rather highlight that reporting should be improved to follow journal guidelines and international stan\-dards---alig\-ning with our own message. 

\subsection{Ethics approval reporting protects people}

Ethics approval processes have been put in place, initially, to ensure that scientists would not cross again dangerous ethical lines \cite{corbie1999continuing,Oak,waggoner2022clinical}. In recent times, however, most of the research that ends up being dangerous and potentially retracted eventually falls on fraudulent practices. Yet, empirical studies of ethics review processes show that review can lead to concrete changes in research procedures and participant-facing materials. Focusing on clinical research, for instance, Stair et al. \cite{stair2001variation} found that IRBs reviewing a multi-center clinical-trial protocol led to the request of changes to study logistics, the research process, and consent forms. With that being said, past research has highlighted that the responses and requested changes from ethics committees are very heterogeneous and rarely consistent \cite{helfand2009variation} but may help identify potentially major concerns with the envisioned research protocol \cite{stark2010variation}, although the detection of flaws in the study design is not something that ethics committees are likely to detect but rather peers through, for instance, registered reports \cite{Chambers:2014:PTG,Nosek_2014,soderberg_2021}. 

What about visualization research? For our field as well, the detection of flaws in the study design is best left to peers, again, ideally before the review process. But are ethics review likely to change our protocol and thus protect participants? One may argue that, considering that the studies are almost always minimal risk, this is unlikely to be the case and they would probably be right. Still, we can report that ethics review has forced us in the past to change one of our study protocol from a full online study of graphic content to an in-person one instead \cite{Reducing}. Whether this change actually protected participant in practice is difficult to quantify, but it has at least made sure to minimize the risks. Nonetheless, we believe that such cases are particularly rare in our community and thus unlikely to happen and we cannot rule out the fact that ethic review of empirical human-subject studies may just as well provide heterogeneous and inconsistent responses (\eg, \cite{kudaibergenova2026decision}). We believe, however, that ethics committees may better enforce data protection and privacy or confidentiality laws (\eg, \cite{angell2010children}) and thus ultimately protect research participants better.

\subsection{Ethics approval reporting protects authors}

\new{The personal story we mentioned in the introduction and share in detail in \autoref{sec:personal}} does not only serve as a footnote to motivate this analysis and the present analysis does not only serve as a justification for our own reporting in the past. We want to highlight here that reporting ethics approval, informed consent, and ethics approval numbers can and should protect authors from potential retractions or unnecessary investigations. Of course, since IEEE TVCG already requires that we mention ethics approval in our manuscripts, we should have been doing it since the journal started publishing our the IEEE VIS proceedings. In practice, this \new{mentioning of ethical approval/consent has largely} not happened but could expose us all to problems and time-consuming investigations and corrections that would make us all, as a field, waste precious research time.

\new{In addition, it is important to consider that corrections and retractions have happened for cases of ethics approval oversight. Of course there are the cases of fraudulent ethics statement or lack of ethics approval that logically lead to retractions (as the one we documented in \autoref{sec:background}), but there are other non-frau\-du\-lent cases that lead to time-con\-su\-ming investigation and need for corrections (\eg \cite{nexus2023correction}). But beyond potential wasted time, we need to consider the potential harm for authors and researchers. While current evidence for the career harms of retractions and corrections would seem to point to the fact that corrections have little impact or could even be positively seen when compared to retractions, the impact on younger scholars can nonetheless remain large with some of them usually leaving after a retraction \cite{Fetterman,memon2025characterizing}. Although the overall findings on this topic differ based on the reason for the correction/retraction (\eg mistake VS misconduct \cite{memon2025characterizing}) and one also has to consider that views on this topic are fast moving considering the recent number of scandals that have plagued academia and the very high rate of retractions in recent years \cite{vannoorden2023retractions}. Lonni's own experience with a correction \cite{frank2023raising} has been that it was positively received by academics and a sign of good faith but this is hardly generalizable evidence and we should probably aim at protecting our colleagues rather than trying to be more data points to answer the questions about the consequences of retracted or corrected science.}

\subsection{\new{Current ethics policies in IEEE VIS and VR}}
\label{sec:VR}

For this year's IEEE VIS conference, we (Tobias together with his OPC co-chairs) have initiated changes during the review process to ensure that the statements are added to accepted papers (also see \autoref{sec:moving-forward}): IPC members were asked to flag those papers where the ethics approval and consent was not sufficient reported, and papers that lacked this information were asked to add it in the revision period.
For IEEE VR 2026, last year, we (Lonni and his program co-chairs) implemented a similar policy \cite{VR_ProgramChairs} with a difference that a failure to report ethics approval would results in a desk reject. This conference was, we believe, the first time any venue in HCI or visualization enforced this so strongly.  In practice, the program chairs decided to be as lenient as possible when in doubt and trusted the authors to report faithfully about the policies of their institution or country. With that being said, however, we noticed an interesting confusion based on the vocabulary around ethics approval. Indeed, it is usually understood that obtaining from an ethics review board a statement that \emph{``the study is exempt from a formal ethics review''} or something similar would be equivalent to obtaining ethics approval. Our IEEE VR form on PCS thus did not include a separate box for ``ethics exemption'' but only one for ethics approval. This box, however, was (mis-)understood by some authors: they thought that not checking it would state that they did not check the ethics situation of their work and they thus reported conflicting information between the PCS checkbox and the free-text form we had set for additional explanation. As such, for IEEE VR 2027, we (the VR program chairs) will pay particular attention to the language used and give additional information and examples on the call for papers and submission guidelines. We (Tobias together with his VIS overall paper co-chairs) will also make corresponding changes to the upcoming submission form for IEEE VIS 2027.

\subsection{Other reporting problems?}

While it was not the initial point of our investigation, as we were inspecting study approvals and study participants we also noticed some interesting aspects that we wish to report here. 

First, anonymization processes can lead to issues. We have found sentences in the finally published papers such as \emph{``The studies have been approved by [Redacted] Institutional Review Board''} or similar sentences in at least two articles. Omission to correct anonymized information (for peer-review) is not rare (as evidenced by two issued corrections outside of our field for similar errors \cite{dasilva2022correction,thompson2026correction}) and is one of the drawbacks of anonymous submissions, in addition to the time that anonymization takes \cite{Snodgrass}. 

Second, some studies involve experts without specifying if they ended-up being co-authors. The practice of involving experts so much that they become co-authors is not particularly shocking in our field. Indeed, in many cases our work with experts involves them at different stages of the research process, during multiple weeks or months (see, \eg, Ye et al. \cite{Ye} reporting on a two-year collaboration) and the experts are directly contributing intellectually to the work being presented even if they become also study participants (before or after being made aware that they would be co-authors) \cite{akbaba2023troubling,Pooryousef,sedlmair2012design}. We would, however, gain from reporting such facts more systematically. This reporting is particularly important for the ethical assessment because it is generally agreed upon that an ethics approval is not needed for the authors of a study themselves (at least outside of clinical research).

Third, we have found that, in some papers, ethics approval language was confusing. Some authors for instance reported that \emph{``participants first provided informed consent under our IRB protocol''}, that \emph{``after giving IRB-approved consent to participate''} the participants started the experiment, or that participant recruitment \emph{``follow[ed] the University’s IRB protocol,''} which are complicated to parse and/or unclear in their actual meaning, and thus difficult to categorize. It is indeed unclear if the IRB protocol mentioned here is the generic one for all studies, that is, a sort of guidelines, or if the authors actually obtained ethics approval for their own study and, if so, whether they received approval for their entire protocol. Similarly, some authors have mentioned that they \emph{``obtained [the participants'] consent to record the study,''} but this statement is unclear whether it indeed cover the fully informed consent for the complete study. Sometimes ethics approval and informed consent are also mixed together, such as by stating \emph{``At the beginning, participants provided informed consent in accordance with our IRB protocol,''} based in which it remains unclear if the whole experiment was indeed approved by the IRB. We strongly encourage authors to be as precise as possible when reporting on ethics approval. 

Fourth, our analysis highlighted that the reporting of ethics approval or consent to participate is not homogeneously reported: some authors mention it in a ``Participants'' subsection, others in ``Procedure'' or ``Tasks,'' while a few papers mentioned it even in the acknowledgments section or only in a figure. Interestingly, some authors mentioned that the ethics documents are available on an OSF repository but we were not always able to locate it. 

One interesting point, however, stemming from our analysis is that LLMs were quite accurate at analyzing this kind of data automatically provided that the number of papers and pages were not too large to lead to a \textit{time-out} of the request.

\subsection{Moving forward}
\label{sec:moving-forward}

Our analysis highlighted that, even in the last two years, most IEEE VIS papers do not report ethics approval or informed consent properly. Only a handful of papers did so, despite the policy in place for the journal. 
\new{As we argued in \autoref{sec:just-not-reported}, many more likely still obtained approval and only failed to report it, even though the venues in which we publish require that we specify this information.}
The TVCG template has already been amended with a section on how to report specific information about ethics approval and it may be even better if this information is separated from the main text (\eg, a separate subsection at the end of the manuscript). 
\new{Our own analysis and that from IHU-MI \cite{frank2023raising}} was particularly time-consuming because this information was difficult to find, and having the it separated from the rest of the manuscript would make such an analysis easier. In an ideal world, ethics approval, their reference numbers, informed consent, participants type (\eg, experts or not, healthy volunteers or sick patients) and numbers \new{would be recorded as} as meta-data so that it is easy to acquire and analyze. We also believe that such \new{reporting} should not count towards the page-limit. In an era in which science is facing the replication crisis and with such emphasis placed on replication, it is perhaps unfortunate that essential information for replicating or faithfully presenting studies is omitted because of a lack of space. While we understand that a page limit also helps to ease the burden on reviewers, we argue that---in the digital age when journals are not often printed any\-more---a page limit is perhaps one of the vestiges of the past we should now move beyond only to consider word limits (as for ACM CHI). 
In a previous paper \cite{frank2023raising} we recommended that editorial processes be amended to allow for the po\-ten\-tial\-ly-con\-fi\-den\-tial ethics approval document to be uploaded. We believe that such an upload could also be implemented for the visualization field but, considering that most of our studies are minimal risk and not clinical research, this may not be necessary. 

One could potentially now suggest that we should all correct our past submission to add the missing information about ethics approval for our stu\-dies---and publishing erratas to papers is certainly possible in TVCG \cite{Luo:2025:DVGE}. For the papers we have analyzed, that would represent \humanResearchLargelyinorrectWord{} (\humanResearchLargelyinorrect) for both years together; \humanResearchLargelyinorrectWordInMMXXIV{} (\humanResearchLargelyinorrectInMMXXIV) in 2024 and \humanResearchLargelyinorrectWordInMMXXV{} (\humanResearchLargelyinorrectInMMXXV) in 2025. In the past, Lonni has argued for the correction of scientific articles to be done as quickly and often as possible \cite{Besancon_correction} and highlighted that we should not feel embarrassed or be punished for having to correct an already published paper. In this case, however, we believe that having to issue corrections for so many articles would be essentially requiring a lot of work from the authors and editors of TVCG for a correction that would not fundamentally change much. It would also further overwhelm research integrity and editing teams that are already drowning under fake and fraudulent papers they have published \cite{abalkina2025stamp,Swart} (yes, even IEEE \cite{abalkina2026openingpandorasboxpaper,besanconarbitrarytimberland}). The easiest way forward is for us to ensure that we do better at reporting ethics in our papers and for the field to start reflecting together on how to do this best. 

\subsection{\new{Recommendations for IEEE VIS/TVCG}}

\new{In summary, in light of our findings and the points we have raised so far in the discussion, we recommend that we should, as a community, implement the following steps:
\begin{enumerate}
    \item \textbf{Do not issue corrections for missing ethics approval statement thus far}, which would waste a lot of our times as authors and editors.
    \item \textbf{Implement strong policies for reporting at submission stage} where editors, ACs, and external reviewers are trained and educated on the topic.
\end{enumerate}}

\new{Furthermore, as a community we could also consider to:
\begin{enumerate}\setcounter{enumi}{2}
    \item \textbf{Have ethics approval reporting standardized}, which would make it more homogeneous, less confusing for readers and reviewers, and easier to find. 
    \item \textbf{Move such standardized reporting outside of the page limit count} as we have done for supplementary materials information and rights in recent years. Ultimately, we also have to remember that IEEE TVCG is not printed and this change would not hurt the journal or its publisher financially.
    \item \textbf{Report more systematically whether expert participants became co-authors}, which would make the dual role of experts as both study participants and contributors explicit and allow readers and reviewers to better assess potential ethical and methodological implications.
\end{enumerate}}


\subsection{Limitations}

Our analysis suffers from obvious limitations. First, we manually coded the papers, each year by a single author (\ie, one year for each of us). Considering that ethics review statements are easy to miss (\eg, at least one of them was only found in a figure) and some distinctions are subjective and personal or open to interpretation, it is impossible for us to guarantee that all of the results presented there would be exactly found by another team doing the same analysis. To limit interpretation biases and errors, however, we implemented the following two steps. First, each author was responsible for a different year and, second, we added the separate LLM analysis with \emph{ChatGPT 5.5 Thinking}. The first step ensures that different subjective flavors find compatible results across years. The second helped us identify cases in which we missed the respective statements. For the VIS 2024 papers, cases where there was a disagreement between the LLM analysis and the coder were resolved on the fly. For the VIS 2025 papers, we highlighted such cases for later analysis and reporting and then verified them by the other author. With this approach and for the 2025 data, we found 14 complete mismatches where the human coder was right and 11 where the LLM was right. Other cases were flagged by the coder or the LLM as ``partial'' or a ``edge-case,'' which we excluded from most of our analysis. 

A second limitation is that we did not manually inspect all of the supplementary material repositories, even though they could have contained ethics approval documents in some cases. Our personal experience reviewing and using such repositories is, however, that they rarely contain these documents. We thus argue that our results are unlikely to change despite this. 

Considering these two limitations, another independent analysis may reveal slightly different numbers---yet we strongly believe that those numbers will be \emph{compatible} with ours and would thus not change any of the take-away messages we presented.

\section{Conclusion}

Visualization research has long depended on human-participant studies, expert feedback, user evaluations, and other forms of engagement with people. Yet, our analysis of IEEE VIS papers from 2024 and 2025 shows that the reporting of ethics approval, ethics approval references, and informed consent remains inconsistent. Only \humanResearchFullyCorrect{} of the \humanResearchOverall{} papers involving human participants reported all three elements \new{for all human involvement}, and many papers did not make it clear whether ethics approval had been obtained, waived, deemed unnecessary, or simply omitted from the manuscript.

This finding should not be read as evidence that VIS research is unethical, nor that the studies we screened lacked appropriate participant protections. In many cases, authors may have obtained approval but not reported it, may have received an exemption, or may have worked under national or institutional frameworks in which formal review was not required for minimal-risk research. The problem is, therefore, primarily one of reporting, not necessarily one of (mis)conduct. However, this reporting matters. It allows reviewers, editors, and ultimately readers to understand how participants were protected as well as helps distinguish genuinely concerning cases from harmless omissions. Moreover, it protects authors from unnecessary or misleading post-publication scrutiny that can waste precious time for many different actors.

We thus argue that ethics reporting should become a routine and visible part of visualization research practice. IEEE VIS and TVCG should provide even clearer guidance on how to report ethics approval, exemption, informed consent, and cases where no formal review was required. They should also consider dedicated manuscript sections or metadata fields for this information, ideally outside the main page limit. Such changes would not only bring VIS papers into closer alignment with publisher requirements, but would also help normalize transparent reporting of human-participant research across the field of visualization.

The goal of this reporting is not to create additional bureaucracy but rather to make our existing practices legible. Better ethics reporting can protect participants, protect authors, support editors, and strengthen the overall credibility of visualization research.

\section*{Supplemental materials}
\label{sec:supplemental_materials}

The goal of our analysis is not to point fingers at authors who have failed to report on ethics. Nonetheless, we still decided to make our data available in the interest of open science, which can be found in our OSF repository at \osfrepo.

\section*{Figure credits and licenses}
\label{sec:figure_credits}

All figures in this paper are our own. They are and remain under our own personal copyright, with the permission to be used here. We also make them available under the \href{https://creativecommons.org/licenses/by/4.0/}{Creative Commons At\-tri\-bu\-tion 4.0 International (\ccLogo\,\ccAttribution\ \mbox{CC BY 4.0})} license at \osfrepo.

\acknowledgments{
We thank, in particular, Inria's COERLE committee for their on-going work to ensure an ethical treatment of work with human participants as well as the teams of Wiley, Monash University, Linköping University, and the COERLE for rigorously but fairly investigating the mentioned accusations.}

\balance
\bibliographystyle{abbrv-doi-hyperref-narrow}
\bibliography{template}

\newpage

\appendix

\section{\new{What counts as ``research involving human subjects''?}}
\label{sec:definition}

\new{We consider \textbf{any work} that contributes \textbf{data or feedback} from \textbf{any human participants other than the authors} of an article itself to count as work that requires an ethical statement. This classification not only includes ``traditional'' ``user studies'' or empirical experiments, but also pilot studies as well a formal or informal feedback from one or more experts etc.}

\section{\new{Personal motivation: From published concerns to shared consequences}}
\label{sec:personal}

\new{In this appendix, we depart} from the impersonal voice of the paper and present \new{a detailed version of the} first-person account from Lonni's perspective\new{, that we summarized in our paper's introduction,} of events that led to consequences on Lonni's work and its ethics and eventually a field-level guideline and reflection.

During the COVID-19 pandemic, I became involved in post-pub\-li\-ca\-tion scrutiny of papers. This process started because I thought I could use some of my skills better to help with the pandemic rather than to solely focus on visualization research. My efforts spanned a wide variety of papers in widely different areas of science but with some focus on epidemiology, with some of my debunking leading to the publication of rebuttals or the retractions of papers that turned out to be wrong (\eg, \cite{meyerowitz2021impact}). I also unearthed, however, problematic practices such as the publication of hundreds of papers with less than 24 hours of peer-reviewing (time spent between submission date and acceptance date), with some of them also having the authors as editors which leads to a rightful questioning of whether the peer-review process was respected \cite{besanccon2021open}. Among some of these papers, one particularly stood out: the study by Gautret et al., which was reviewed in less 24 hours (although editorial evidence of this has, since then, been removed \cite{barraud2023article}) which also contained problematic methodology and ethics (see \cite{barraud2023article}). The study was the leading argument to promote hydroxychloroquine and azithromycin for COVID-19, which then became massively covered in the news. Today, we have strong evidence that the proposed treatment does not work and, after four years of a difficult battle, the initial paper has been retracted \cite{gautret2020retracted,ogrady2024infamous,van2024controversial}. 
The paper also raised questions about research ethics oversight, including whether the approval described in the article corresponded to the study that had actually been conducted by the authors.

This question led to a broader investigation. Together with colleagues, I examined a large set of publications affiliated with the IHU Mediterranee Infection (IHU-MI) to determine whether the ethics approval statements reported in their articles were sufficient to understand the oversight under which the studies had been conducted. We found that, in many cases, the statements were missing, unclear, reused across apparently different studies, or difficult to reconcile with the procedures, samples, participants, or countries described in the articles \cite{frank2023raising}. In this context, ethics approval statements were not peripheral administrative details. They were often the only public information available to evaluate whether research involving human participants had been conducted under appropriate ethical oversight. The most astonishing finding was that the same ethics approval statement and number had been reused, back then, 248 times (now more than 250 times) for studies that were conducted in different countries or on different populations or with different clinical samples being taken. Following the publication or our article, the results became media-worthy and warranted a full investigation and reporting from both Science \cite{ogrady2024reckoning} and Nature \cite{ro2025covid} and the amendment of the Helsinki declaration to prevent such abuse in the future. Our article could not prove directly that any misconduct happened, but it raised several questions that editors were then tasked to investigate. To date and following these investigations, 72 papers have been retracted and 268 have received editorial expressions of concerns,\footnote{Source: \href{https://ihu-correction.com}{\texttt{ihu\discretionary{}{-}{-}correction\discretionary{}{.}{.}com}}.} while we have just published our latest analysis showing ethical or legal concerns in 853 articles of the same institute \cite{frank2026followup}

After this work was published and publicized, similar scrutiny was directed at some of my own publications, including visualization and human-computer interaction papers co-authored with Tobias. Complaints were sent to publishers (only Wiley as far as we know) and universities (Inria, Monash University, and Linköping University as far as we know) asking them to investigate, and in some cases retract, these papers on the grounds that they did not explicitly include ethics approval statements that the journals required.
These requests were brought forward despite the fact that, for the papers concerned, the relevant approvals or participant-protection procedures either did no exist or had been followed. When I conducted my PhD work at Inria, institutional guidelines did not require ethics approval for the type of non-clinical HCI and visualization studies I was conducting. In some cases, I nevertheless sought such approval. In others, I at least followed the participant-protection practices expected in the field, including informed consent and anonymization. What I did not always do was report these procedures explicitly in the paper, which was consistent with common reporting practices in visualization at the time.
Together, we made this argument in our response to Wiley, the publisher of the EuroVis proceedings and the first publisher to escalate these concerns to us. To support it, we also provided Wiley with an analysis of recent EuroVis papers, showing that the absence of explicit ethics approval statements was not unusual in the field.

I want to strongly clarify that I am very happy that post-publication scrutiny is given to my papers and that publishers take it seriously. I have, in fact, argued for stronger and more careful scrutiny (pre- and post-publication) and better as well as faster follow-up from publishers \cite{Besanccon2026conference,Besancon_correction}. Still, this reversal of scrutiny was personally and professionally taxing, because justifying the ethics of published work is a complicated matter that involves being able to explain national regulations which are rarely available in English outside of English-speaking countries. This situation was also problematic because the concerns about my work were first raised on PubPeer (a platform for post publication peer-review \cite{barbour2020pubpeer}) and on social media to suggest that I had to follow clinical guidelines which was also nonsensical and took a lot of time to rebut despite how misguided those were.\footnote{E.\,g., \href{https://pubpeer.com/publications/8A7754CD25507278766E728C0E1B33}{\textls[-35]{\texttt{pubpeer\discretionary{}{.}{.}com\discretionary{/}{}{/}publications\discretionary{/}{}{/}8A7754CD25507278766E728C0\discretionary{E}{}{E}1B33}}}.} Yet, the comments and my own investigation then revealed something about our practices in the field: visualization and HCI papers often reported the design of user studies in detail, while saying little about ethics approval, exemption, or informed consent. This omission does not necessarily mean that the studies were unethical or not reviewed, but, I believe, reflects the reporting norms of the field, the minimal-risk nature of many studies, or local institutional procedures that did not map neatly onto clinical models of ethics approval.

From my own and Tobias' reviewing and paper reading experiences, it appeared that the norms of the field that we mentioned in our response to Wiley (whose investigation cleared us) might also apply to IEEE VIS and TVCG. 

\section{\new{Ethics guidelines from the various publishers relevant for visualization}}
\label{sec:regulations}

\new{Below we quote the relevant stated regulations for the publishers of those journals (and conferences), in which we in visualization most frequently publish.}

\subsection{\new{IEEE (Transactions on Visualization and Computer Graphics and many conferences)}}

\new{Excerpt from the \href{https://pspb.ieee.org/images/files/PSPB/opsmanual.pdf}{IEEE Publication Services and Products Board Operations Manual}, version of January 2026, Section 8.1.1.E:}

\begin{quote}
\new{\emph{Authors of articles reporting on research involving human subjects or animals, including but extending beyond medical research, shall include a statement in the article that the research was performed under the oversight of an institutional review board or equivalent local/regional body, including the official name of the IRB/ethics committee, or include an explanation as to why such a review was not conducted. For research involving human subjects, authors shall also report that consent from the human subjects in the research was obtained or explain why consent was not obtained.}}
\end{quote}

\subsection{\new{Wiley (Computer Graphics Forum)}}
\label{sec:regulations:wiley}

\new{Excerpt from \href{https://authors.wiley.com/ethics-guidelines/index.html}{\texttt{authors\discretionary{}{.}{.}wiley\discretionary{}{.}{.}com\discretionary{/}{}{/}ethics\discretionary{}{-}{-}guidelines\discretionary{/}{}{/}index\discretionary{}{.}{.}html}}, as of July 2026:}

\begin{quote}
\new{\emph{Authors must confirm in their manuscript that ethical approval was received before the beginning of the study. The name of the approving ethics committee or institutional review board must also be included, along with any associated approval numbers. Authors are encouraged to follow the \href{https://www.care-statement.org/}{CARE guidelines} when preparing case reports. Ethics approval must be obtained before a study begins; retrospective ethics approval is generally not accepted. Editors may request additional information or supporting documentation for studies that lack prospective ethics approval, and they may decline to consider such manuscripts.}}
\end{quote}

\subsection{\new{Elsevier (Computers \& Graphics)}}

\new{Excerpt from \href{https://www.elsevier.com/about/policies-and-standards/research-ethics\#0-human-research}{\texttt{elsevier\discretionary{}{.}{.}com\discretionary{/}{}{/}about\discretionary{/}{}{/}policies\discretionary{}{-}{-}and\discretionary{}{-}{-}sta\discretionary{n}{}{n}dards\discretionary{/}{}{/}research\discretionary{}{-}{-}ethics\#0\discretionary{}{-}{-}human\discretionary{}{-}{-}research}}, as of July 2026:}

\begin{quote}
\new{\emph{If the work involves human participants, living or deceased, including their data and/or biological materials (e.g. organs, tissues), the author should ensure that all studies were performed in compliance with relevant laws, regulations, and institutional guidelines and that the appropriate institutional committee(s) has/have approved them. The authors should include a statement in the manuscript that confirms these requirements have been met. The statement should contain the date and reference number of the ethical approval(s) obtained (for exempt studies, see below).}} 

\new{\emph{Research involving human participants should be carried out in accordance with the \href{https://www.wma.net/policies-post/wma-declaration-of-helsinki/}{World Medical Association Declaration of Helsinki: Ethical Principles for Medical Research Involving Human Participants} and any other guidelines that are specific to the relevant discipline(s) and type(s) of research. [...]}}

\new{\emph{Informed consent must be obtained for research with human participants, including their data or biological materials. Consent must be given without coercion, and the privacy rights of human participants must always be observed. Authors should include a statement in the manuscript confirming that they obtained informed consent from all participants or their parent, guardian, or other individual with legal authority to act on their behalf (“Legal Representative”) where applicable. Consent for publication should also be obtained.}}

\new{\emph{Appropriate written consents, permissions and releases must be obtained where an author wishes to include case details, videos, recordings, images, photographs, and illustrations (or any other identifiable form) of patients and any other individuals in an Elsevier publication (see “Consent requirements for case reports, personal information, and images” below). Written consents must be retained by the author and copies of the consents or evidence that such consents have been obtained must be provided to Elsevier on request.}}

\new{\emph{For research with human participants where only verbal informed consent was obtained, authors should include justification for the absence of written consent, the name of the approving ethics committee, and a statement on how verbal consent was recorded. Authors should be aware that written (or electronic) consent is required for certain studies under national laws, including but not limited to clinical trials, case studies, studies with vulnerable populations, and studies that contain any personally identifiable data.}}
\end{quote}

\subsection{\new{ACM (CHI etc.)}}

\new{Excerpt from the \href{https://www.acm.org/publications/policies/research-involving-human-participants-and-subjects}{2021 ACM Publications policy on research involving humans}, as of July 2026:}

\begin{quote}
\new{\emph{All authors conducting research involving human participants and subjects must meet appropriate ethical and legal standards guiding such research. In particular, ACM authors must ensure that their human research planning, conduct, and reporting are consistent with their local governing laws and regulations and the general principles detailed below. It is important to note that something may not be ethical even though it is not prohibited by local law or regulation, and thus authors should also ensure alignment that their research practices are compliant with the \href{https://www.acm.org/code-of-ethics}{ACM Code of Ethics and Professional Conduct} and international and national standards for such research, such as \href{https://www.wma.net/policies-post/wma-declaration-of-helsinki-ethical-principles-for-medical-research-involving-human-subjects/}{The Declaration of Helsinki}, \href{https://www.hhs.gov/ohrp/regulations-and-policy/belmont-report/index.html}{The Belmont Report}, and \href{https://www.hhs.gov/ohrp/regulations-and-policy/regulations/finalized-revisions-common-rule/index.html}{The Common Rule}, including, but not limited to:}
\begin{itemize}
\item \emph{minimization of potential harms, making sure any risks are justified by potential benefits}
\item \emph{protection for the privacy and right to self-determination of participants and subjects}
\item \emph{adhering to relevant institutional, local, national, and international regulations}
\item \emph{adhering to the principle of informed consent}
\item \emph{adhering to the principle of justice}
\item \emph{adherence with all other applicable ACM policies}
\end{itemize}}

\new{\textls[-7]{\emph{Where such research is conducted in countries where no such local governing laws and regulations related to human participant and subject research exist, Authors must at a bare minimum be prepared to show compliance with the above detailed principles. Authors should also declare any potential conflicts of interest in compliance with the \href{https://www.acm.org/publications/policies/conflict-of-interest}{Conflict of Interest Policy for ACM Publications}, so that reviewers and editors may determine whether the declared COIs are significant enough to warrant rejection of the Work or another appropriate remedy. [...]}}}

\new{\emph{It is the authors’ responsibility (each author individually and the authors collectively) to comply with and provide evidence of compliance with this Policy.  Where local ethical review boards are required, authors are responsible for having their research reviewed and approved by such boards.  Authors are also responsible for the overall ethical conduct of their research.  All ACM Authors must be prepared to provide documentary evidence to ACM that they have adhered to local ethical and legal standards, as ACM may require documentary evidence of such approval at any time following submission of the Work and prior to or after publication of the Work.}}
\end{quote}

\subsection{\new{Eurographics (several conferences and symposia)}}

\new{Other than for the papers that are published---either as regular articles or as conference papers in special issues---by Wiley in the EG's signature journal, Computer Graphics Forum (see \autoref{sec:regulations:wiley}), the Eurographics Association does not (yet---as of Summer 2026) regulate how to report ethical approval or informed consent for work with human participants.}

\end{document}